\documentstyle[twocolumn,prl,aps,epsf]{revtex}

\title{On Anisotropic Transport in High Landau Levels.}

\begin{document}
\newcommand{\showfigures}{yes}     
\maketitle
Anisotropy of resistance in two orthogonal directions has been observed since the development in 1987 of high mobility $\delta$-doped GaAs heterostructures.[1] The phenomenon was brought forward in a talk by Stormer $\it et\ al.$[2], and by Lilly $\it et\ al.$ and Du $\it et\ al.$[3] The anisotropy is observed both at zero magnetic field $B=0$ and at quantizing $B$, and is particularly strong in an intermediate range of $B$ corresponding to occupation of several ($N \sim 3$) Landau levels. The anisotropy was attributed to possible formation of charge density waves (CDW), specifically "stripe" and/or "bubble" phases [4] below temperature $T \sim1$ K.

Here we would like to point out that: (i) according to the edge-bulk transport models for the quantum Hall regime [5], the direction of low bulk conductance is actually perpendicular to that expected naively; and (ii) the values of experimental "high resistance" peaks correspond to value of bulk conductance $ \approx 1 e^{2}/h$.

Using the convention of [3], we chose $X$-axis parallel to the direction of high four-terminal resistance, so that $R _{xx} > R _{yy}$. Then, using the approximation good for a long sample, we can express the two measured four-terminal resistances in terms of the local bulk conductivities $\sigma _{xx}$ and $\sigma _{yy}$: 

\begin{equation} 
R _{xx} \approx \Sigma _{yy}/(\sigma _{xy})^2$ and  $R _{yy} \approx  \Sigma _{xx}/(\sigma _{xy})^2,
\label{A}
\end{equation} 

\noindent where bulk conductance $ \Sigma _{xx} \approx \Box \sigma _{xx}$, and  $\Box \sim 1$ is "number of squares". It is clear from these expressions that  $R _{xx} \gg R _{yy}$ corresponds to  $\sigma _{xx} \ll \sigma _{yy}$; that is, high measured resistance in one direction ($X$) corresponds to high bulk conductivity in the orthogonal direction ($Y$).

A moderate in-plane magnetic field $B _{\|} $ is expected to mix the $X$ and $Y$ conduction, and thus to reduce the ratio $R _{xx}/R _{yy} = \Sigma _{yy}/ \Sigma _{xx}$.  If the direction of the in-plane field is perpendicular to the direction $X$ of lower bulk conductivity the effect of admixture of higher conductivity is stronger, while the effect of mixing is weaker for the direction of  $B _{\|} $ parallel to the direction of lower bulk conductivity.  We would thus expect the effect of the in-plane field be more pronounced for $B _{\|} $ in the $Y$ direction because  $\sigma _{xx} \ll \sigma _{yy}$, consistent with the data reported in the preprints of Ref. 3. This analysis obviously does not apply to strong $B _{\|} $, where spin polarization and subband mixing may lead to a phase transition.

Further, inverting Eq.(\ref{A}) for the low $T$ saturated values of $R _{xx}$,[3] quite remarkably, we obtain the peak values of conductance  $\Sigma _{yy} \approx 1 e^{2}/h$ within 20 percent for all $R _{xx}$ peaks ($\nu = 9/2$ to $\nu = 19/2$) for all samples.[6] This value seems to be systematically different from  $\sigma _{xx} \approx \frac 1 2 e^{2}/h$ expected for the isotropic peak bulk conductivity in the integer QH regime, and observed in Corbino geometry samples [7]. We also note that the peak values of  $R _{yy}$ correspond to lower conductance $\Sigma _{xx} \sim 0.1 e^{2}/h $; it is much more variable, both for different peaks in a given sample and from sample to sample.[3]

Interpretation of the anisotropic transport experiments [2,3] in terms of formation of a pinned CDW then requires CDW to be pinned in the $X$ direction (at $B _{\|} = 0$), the direction of lower bulk conductance. Such interpretation then also implies that the conduction by this CDW in $Y$ direction is coherent, as in a spin-polarized one "channel", resulting in  $\Sigma _{yy} \approx 1 e^{2}/h$, and that the pinning is suppressed by a perpendicular in-plane magnetic field.

The temperature dependence of anisotropic $R _{xx}$ reported in [3] is not much different from that reported for the nonlocal resistance $R _{NL} \propto \beta T/[$exp$(\beta T) - 1]$, where $\beta = (80$ mK$)^{-1}$ (second paper of [5], where the temperature dependence was attributed to the bulk-edge scattering length decreasing at higher $T$).  While the similarity of the two temperature dependencies may be accidental, and certainly does not explain the observed anisotropy, nonetheless this fact suggests that a quantitative analysis of the experimental data should include analysis in terms of bulk-edge transport models.

\vspace{12pt}
\noindent V. J. Goldman\\
Department of Physics\\
SUNY at Stony Brook\\
Stony Brook, NY 11794-3400\\
\\
1. J. H. English {\em et al.}, Appl. Phys. Lett. {\bf 50}, 1826 (1987);  M. Shayegan $\it et\ al.$, Appl. Phys. Lett. {\bf 52}, 1086 (1988). \\
2. H. L. Stormer {\em et al.}, Bull. Am. Phys. Soc. {\bf 38}, 235 (1993).\\
3. M. P. Lilly {\em et al.}, Phys. Rev. Lett. {\bf 82}, 394 (1999); R. R. Du {\em et al.}, Solid State Commun. {\bf 109}, 389 (1999); and preprints.\\ 
4. M. M. Fogler, A. A. Koulakov, and B. I. Shklovskii, Phys. Rev. B {\bf 54}, 1853 (1996); R. Moessner and J. T. Chalker, Phys. Rev. B {\bf 54}, 5006 (1996). \\
5. See, e. g., J. K. Wang and V. J. Goldman, Phys. Rev. Lett. {\bf 67}, 749 (1991);  Phys. Rev. B {\bf 45}, 13 479 (1992). \\
6. $\Sigma _{yy} = 1 e^{2}/h$ corresponds to ($R _{xx}$ at $\nu$): 1220 $\Omega$ at $9/2$; 830 $\Omega$ at $11/2$; 600 $\Omega$ at $13/2$; 450 $\Omega$ at $15/2$; 350 $\Omega$ at $17/2$; 280 $\Omega$ at $19/2$, $\em etc.$  \\
7.  D. E. Khmelnitskii, JETP Lett {\bf 38}, 552 (1983); L. P. Rokhinson {\em et al.}, Solid State Commun. {\bf 96}, 309 (1995).\\ 

\end{document}